\documentclass{aip-cp}

\usepackage[numbers]{natbib}
\usepackage{rotating}
\usepackage{graphicx}
\usepackage{savesym}
\savesymbol{iint}
\savesymbol{iiint}
\savesymbol{iiiint}
\savesymbol{idotsint}
\usepackage{amsmath}
\usepackage{amsfonts}
\usepackage{empheq}
\usepackage{amssymb}
\usepackage{enumitem}
\usepackage{subfigure}
\usepackage{textcomp}
\usepackage[nottoc]{tocbibind}

\begin{document}

\title{HAWC Upgrade for Multi-TeV $\gamma$-ray Detection}

\author[aff1]{A. Jardin-Blicq\corref{cor1}}
\author[aff1]{V. Joshi}
\eaddress{vikas.joshi@mpi-hd.mpg.de}

\author[aff2]{for the HAWC collaboration}
\affil[aff1]{Max Planck Institut f\"{u}r Kernphysik, Heidelberg}
\affil[aff2]{For a complete author list, see www.hawc-observatory.org/collaboration/}
\corresp[cor1]{armelle.jardin-blicq@mpi-hd.mpg.de}

\maketitle

\begin{abstract}
The High Altitude Water Cherenkov (HAWC) high-energy $\gamma$-ray observatory was completed in march 2015 in central Mexico. The detector, consisting of 300 
water tanks, is currently being upgraded to improve its performance at Multi-TeV energies, with a sparse array of small water Cherenkov tanks. It will extend 
the instrumental area by a factor of 4, and enhance the sensitivity at the highest energies.
In this contribution, the current status of the observatory is presented, as well as the coming upgrade. The electronics and the readout system for the new sparse 
array of small water tanks are also desrcibed, and results from simulations performed to optimize the performance of the array are discussed.
\end{abstract}

\section{INTRODUCTION}
The 300 water tanks composing HAWC are covering an area of 22000 m$^2$, each tank containing 200 kl of purified water and four photomultiplier tubes (PMT). HAWC records events in the 0.1-100 TeV energy range. 
It has now completed its first year of operation, and the first results are already showing an improvement in sensitivity to TeV $\gamma$-ray sources an order of magnitude better than the 
previous generation of water Cherenkov detectors. The first catalog of sources is being prepared, with several of them 
observed for the first time in $\gamma$-rays, with no known counterpart.
However, HAWC encounters difficulties 
reconstructing showers whose core falls outside the array. A sparse outrigger array of small water Cherenkov tanks aims to increase the present fraction of 
well-reconstructed showers at multi-TeV energies by a factor of 3 to 4 by improving the accuracy of their core position determination. Such an outrigger array 
would consist of 350 small water Cherenkov tanks of 2.5 m$^3$, each containing one PMT, distributed over an area four times larger than HAWC.

\section{MOTIVATION FOR THE UPGRADE}
When a highly energetic particle enters the atmosphere and interacts with air nuclei, it creates a shower of secondary particles that eventually reach the ground, 
leaving a "footprint" on the detector. Above $\sim$10 TeV, the shower footprint on the ground becomes comparable to the size of the array. Therefore, most of the showers above 10 TeV 
are not contained in the array. In particular, when the core of the shower is not inside the array, the shower properties, namely the core location and the incident angle, cannot be 
correctly reconstructed. Indeed, from 
the main array point of view, a high energy shower whose core falls outside the array is hard to distinguish from a low energy one with a core falling on the edge 
of the array. The misinterpretation of high energy $\gamma$-ray showers reduces HAWC performance at these energies. 
This is the main motivation for the deployment of 350 small water tanks of  2.5 m$^3$ called “outriggers” around the main array \cite{2015arXiv150904269S}. They will be separated by 
a distance of 12 to 18 m. Each of them will contain one Hamamatsu R5912 8 inch PMT. The outrigger array will allow us to :
\begin{enumerate}
 \item Accurately determine the core position, and therefore the direction and the  energy of the shower
 \item Increase the effective area above 10 TeV by a factor of 3-4 and hence enhance the sensitivity above this energy
\end{enumerate}


\section{HARDWARE : TRIGGER AND READOUT}
\begin{figure}[h]
    \begin{minipage}[ht]{0.48\linewidth}
       \centering 
    \includegraphics[width=223pt]{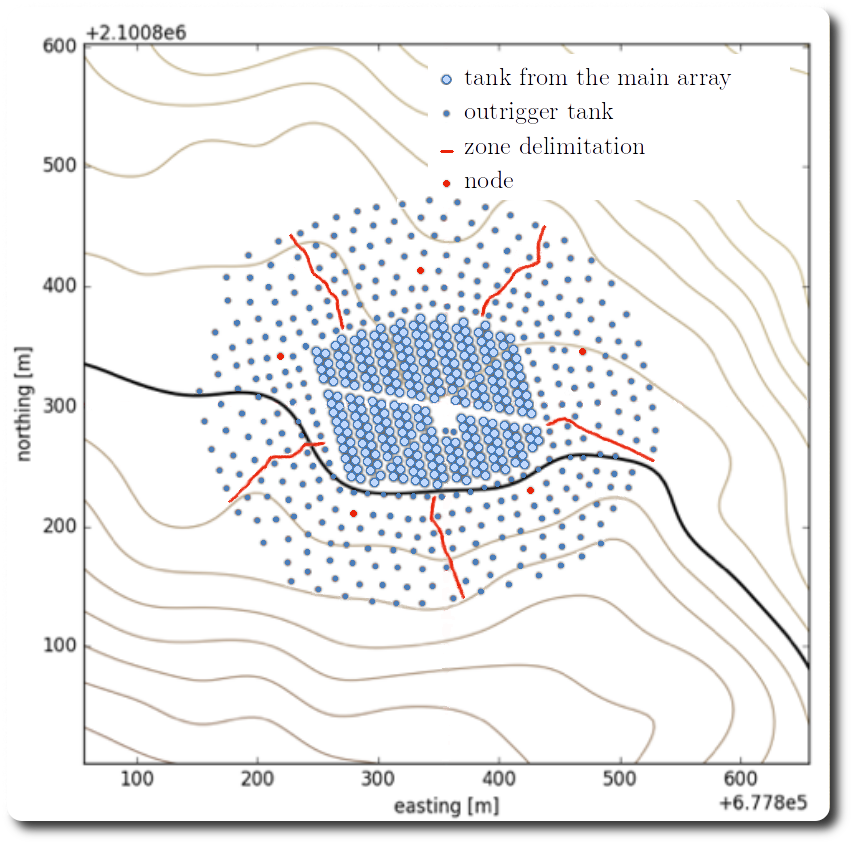}
 \caption{The outrigger layout surrounding the HAWC main array.}
    \end{minipage}\hfill
    \begin{minipage}[ht]{0.48\linewidth}   
       \centering 
    \includegraphics[width=225pt]{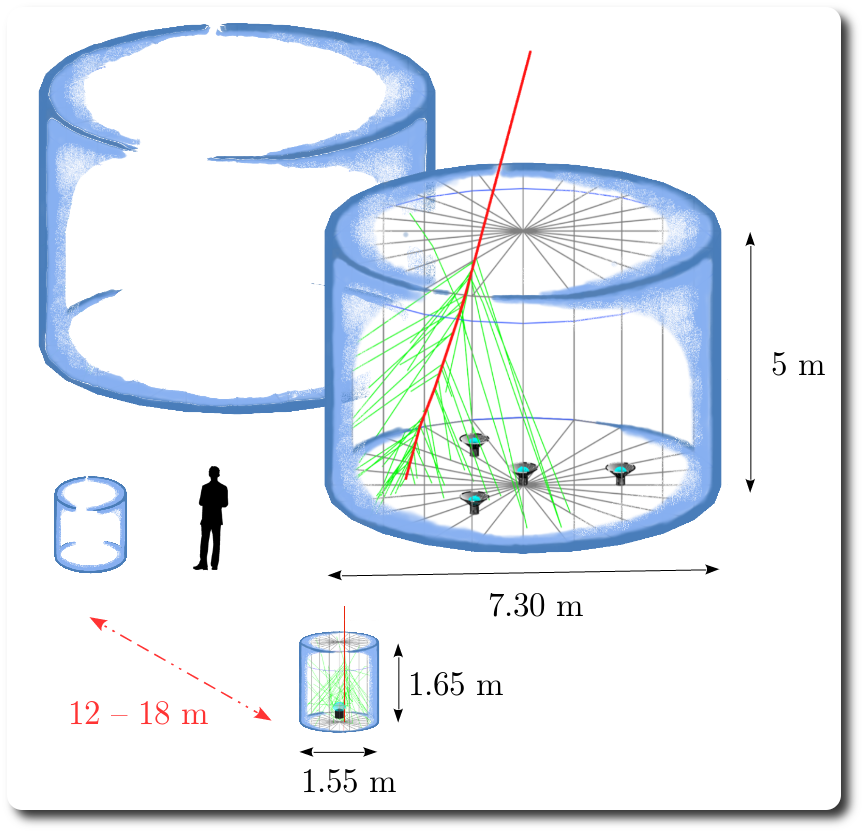}
 \caption{Schematics of a tank from the main array compared to an outrigger tank.}
    \end{minipage}
     \caption{Left panel : the outrigger layout surrounding the HAWC main array divided into 5 sections. The signals of the PMTs in each section are collected in the corresponding node (red dot). 
     Right panel : schematic of a tank from the main array (top) compared to an outrigger tank (bottom).}
     \label{tanks2}
\end{figure}

The outrigger array is very similar to the camera of an Imaging Atmospheric Cherenkov Telescope, each PMT in each outrigger being a pixel of the camera. 
The HAWC $\gamma$-ray observatory is currently taking advantage of the readout and trigger electronics that is being developed for FlashCAM \cite{2013arXiv1307.3677P}, 
a proposed camera solution for a mid-size telescope of the future Cherenkov Telescope Array (CTA).
Hence, it will be used and adapted to the outrigger array :  each of the outrigger PMT will be read out with the Flash-ADC board developed for FlashCAM.
It has been named Flash Adc eLectronics for the Cherenkov Outrigger Node (FALCON).
The sparse outrigger array will be divided into 5 sections with $\sim$70 small water tanks. One node is settled in each of them collecting the signals from the corresponding 
outrigger tanks as shown Fig. \ref{tanks2}. Each node has a crate containing 3 Flash-ADC boards, and each of them can digitize 24 channels with a sampling speed of 250 MHz with a 12 bit accuracy. 
It also allows for a flexible digital multiplicity trigger. Full waveforms, with settable length (typically 40 samples i.e. 160 ns), are read out and processed further for charge extraction 
and signal timing.

\section{SIMULATIONS}
In order to optimise the outrigger array, extensive simulations are being performed.
The lateral amplitude distribution of the shower is fitted using a likelihood fit method in order to constrain : 
\begin{itemize}
 \item the core location
 \item the shower energy
 \item the depth of the shower maximum
\end{itemize}

\begin{figure}[h!]
  \centerline{\includegraphics[width=300pt]{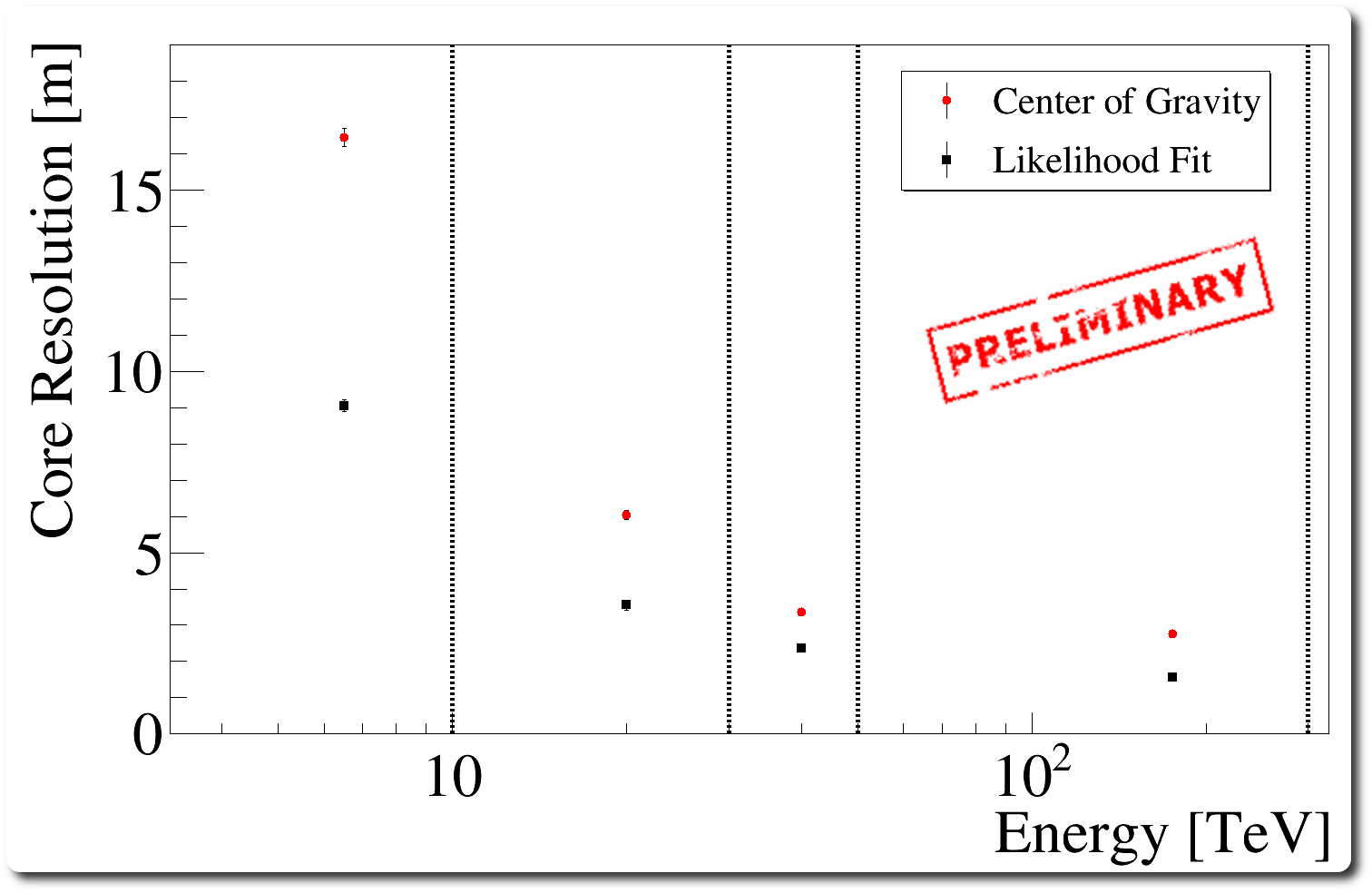}}
  \caption{The core resolution obtained with a likelihood fit, compared with a method where the core is defined as the centre of gravity of the signal}
  \label{core_resolution_plot}
\end{figure}

On Fig. \ref{core_resolution_plot}, it can be seen that, using this method, a core resolution of 2 to 3 m can be achieved by using only an array of sparse outriggers, at energies 
above 10 TeV for a 0\textdegree \ zenith angle. For comparision, the core resolution using only the main array is of the order of one meter. The next step is to merge the likelihood 
fit method for the outrigger array with the one for the main array to quantify the improvement in the core resolution for the highest energy showers. 


\section{TOWARDS THE OUTRIGGER ARRAY}
The outrigger array is currently being deployed : the first outrigger prototypes are taking data on the site, with the FALCON electronics. After thoroughly testing, it will then be 
integrated to the central DAQ. The sections will then be deployed one by one, and the first one is planned to be completed by the beginnig of next year. The full outrigger 
array will be completed next year.


\section{ACKNOWLEDGMENTS}
We acknowledge the support from: the US National Science Foundation (NSF); the US Department of Energy
Office of High-Energy Physics; the Laboratory Directed Research and Development (LDRD) program of Los Alamos
National Laboratory; Consejo Nacional de Ciencia y Tecnología (CONACyT), Mexico (grants 260378, 55155, 105666,
122331, 132197, 167281); Red de Física de Altas Energías, Mexico; DGAPA-UNAM (grants IG100414-3, IN108713,
IN121309, IN115409, IN113612); VIEP-BUAP (grant 161-EXC-2011); the University of Wisconsin Alumni Research
Foundation; the Institute of Geophysics, Planetary Physics, and Signatures at Los Alamos National Laboratory; the Luc
Binette Foundation UNAM Postdoctoral Fellowship program.

\nocite{*}
\bibliographystyle{aipnum-cp}%
\bibliography{sample}%

\begin{thebibliography}{2}%
\makeatletter
\providecommand \@ifxundefined [1]{%
 \@ifx{#1\undefined}
}%
\providecommand \@ifnum [1]{%
 \ifnum #1\expandafter \@firstoftwo
 \else \expandafter \@secondoftwo
 \fi
}%
\providecommand \@ifx [1]{%
 \ifx #1\expandafter \@firstoftwo
 \else \expandafter \@secondoftwo
 \fi
}%
\providecommand \natexlab [1]{#1}%
\providecommand \enquote  [1]{``#1''}%
\providecommand \bibnamefont  [1]{#1}%
\providecommand \bibfnamefont [1]{#1}%
\providecommand \citenamefont [1]{#1}%
\providecommand \href@noop [0]{\@secondoftwo}%
\providecommand \href [0]{\begingroup \@sanitize@url \@href}%
\providecommand \@href[1]{\@@startlink{#1}\@@href}%
\providecommand \@@href[1]{\endgroup#1\@@endlink}%
\providecommand \@sanitize@url [0]{\catcode `\$12\catcode `\&12\catcode
  `\#12\catcode `\^12\catcode `\_12\catcode `\%12\relax}%
\providecommand \@@startlink[1]{}%
\providecommand \@@endlink[0]{}%
\providecommand \url  [0]{\begingroup\@sanitize@url \@url }%
\providecommand \@url [1]{\endgroup\@href {#1}{\urlprefix }}%
\providecommand \urlprefix  [0]{URL }%
\providecommand \Eprint [0]{\href }%
\providecommand \doibase [0]{http://dx.doi.org/}%
\providecommand \selectlanguage [0]{\@gobble}%
\providecommand \bibinfo  [0]{\@secondoftwo}%
\providecommand \bibfield  [0]{\@secondoftwo}%
\providecommand \translation [1]{[#1]}%
\providecommand \BibitemOpen [0]{}%
\providecommand \bibitemStop [0]{}%
\providecommand \bibitemNoStop [0]{.\EOS\space}%
\providecommand \EOS [0]{\spacefactor3000\relax}%
\providecommand \BibitemShut  [1]{\csname bibitem#1\endcsname}%
\let\auto@bib@innerbib\@empty
\bibitem [{\citenamefont {{Sandoval}}(2015)}]{2015arXiv150904269S}%
  \BibitemOpen
  \bibfield  {author} {\bibinfo {author} {\bibfnamefont {A.}~\bibnamefont
  {{Sandoval}}},\ }\bibfield  {booktitle} {\emph {\bibinfo {booktitle} {Proc.
  of the 34rd ICRC, La Hague}},\ }\href@noop {} {\ September (\bibinfo {year}
  {2015})},\ \Eprint {http://arxiv.org/abs/1509.04269} {astro-ph.IM:1509.04269}
  \BibitemShut {NoStop}%
\bibitem [{\citenamefont {{P{\"u}hlhofer}}\ \emph {et~al.}(2013)\citenamefont
  {{P{\"u}hlhofer}}, \citenamefont {{Bauer}}, \citenamefont {{Eisenkolb}},
  \citenamefont {{Florin}}, \citenamefont {{F{\"o}hr}}, \citenamefont
  {{Gadola}}, \citenamefont {{Hermann}}, \citenamefont {{Kalkuhl}},
  \citenamefont {{Kasperek}}, \citenamefont {{Kihm}}, \citenamefont {{Koziol}},
  \citenamefont {{Manalaysay}}, \citenamefont {{Marszalek}}, \citenamefont
  {{Rajda}}, \citenamefont {{Romaszkan}}, \citenamefont {{Rupinski}},
  \citenamefont {{Schanz}}, \citenamefont {{Steiner}}, \citenamefont
  {{Straumann}}, \citenamefont {{Tenzer}}, \citenamefont {{Vollhardt}},
  \citenamefont {{Weitzel}}, \citenamefont {{Winiarski}}, \citenamefont
  {{Zietara}},\ and\ \citenamefont {{CTA consortium}}}]{2013arXiv1307.3677P}%
  \BibitemOpen
  \bibfield  {author} {\bibinfo {author} {\bibfnamefont {G.}~\bibnamefont
  {{P{\"u}hlhofer}}}, \bibinfo {author} {\bibfnamefont {C.}~\bibnamefont
  {{Bauer}}}, \bibinfo {author} {\bibfnamefont {F.}~\bibnamefont
  {{Eisenkolb}}}, \bibinfo {author} {\bibfnamefont {D.}~\bibnamefont
  {{Florin}}}, \bibinfo {author} {\bibfnamefont {C.}~\bibnamefont
  {{F{\"o}hr}}}, \bibinfo {author} {\bibfnamefont {A.}~\bibnamefont
  {{Gadola}}}, \bibinfo {author} {\bibfnamefont {G.}~\bibnamefont {{Hermann}}},
  \bibinfo {author} {\bibfnamefont {C.}~\bibnamefont {{Kalkuhl}}}, \bibinfo
  {author} {\bibfnamefont {J.}~\bibnamefont {{Kasperek}}}, \bibinfo {author}
  {\bibfnamefont {T.}~\bibnamefont {{Kihm}}}, \bibinfo {author} {\bibfnamefont
  {J.}~\bibnamefont {{Koziol}}}, \bibinfo {author} {\bibfnamefont
  {A.}~\bibnamefont {{Manalaysay}}}, \bibinfo {author} {\bibfnamefont
  {A.}~\bibnamefont {{Marszalek}}}, \bibinfo {author} {\bibfnamefont {P.~J.}\
  \bibnamefont {{Rajda}}}, \bibinfo {author} {\bibfnamefont {W.}~\bibnamefont
  {{Romaszkan}}}, \bibinfo {author} {\bibfnamefont {M.}~\bibnamefont
  {{Rupinski}}}, \bibinfo {author} {\bibfnamefont {T.}~\bibnamefont
  {{Schanz}}}, \bibinfo {author} {\bibfnamefont {S.}~\bibnamefont {{Steiner}}},
  \bibinfo {author} {\bibfnamefont {U.}~\bibnamefont {{Straumann}}}, \bibinfo
  {author} {\bibfnamefont {C.}~\bibnamefont {{Tenzer}}}, \bibinfo {author}
  {\bibfnamefont {A.}~\bibnamefont {{Vollhardt}}}, \bibinfo {author}
  {\bibfnamefont {Q.}~\bibnamefont {{Weitzel}}}, \bibinfo {author}
  {\bibfnamefont {K.}~\bibnamefont {{Winiarski}}}, \bibinfo {author}
  {\bibfnamefont {K.}~\bibnamefont {{Zietara}}}, \ and\ \bibinfo {author}
  {\bibfnamefont {f.~t.}\ \bibnamefont {{CTA consortium}}},\ }\bibfield
  {booktitle} {\emph {\bibinfo {booktitle} {Proc. of the 33rd ICRC, Rio de
  Janeiro}},\ }\href@noop {} {\ July (\bibinfo {year} {2013})},\ \Eprint
  {http://arxiv.org/abs/1307.3677} {astro-ph.IM:1307.3677} \BibitemShut
  {NoStop}%
\end{thebibliography}%

\end{document}